\documentclass{elsart5p}

\usepackage{graphicx,lineno}

\usepackage{amssymb}

\begin{document}
\linenumbers
\begin{frontmatter}



\title{STUDY OF GAS PURIFIERS FOR THE CMS RPC DETECTOR}


\author[lnf]{L. Benussi}
\author[lnf]{S. Bianco}
\author[lnf,cern,sapienza]{S.Colafranceschi}
\author[lnf]{F.L. Fabbri}
\author[lnf,sapienza]{F. Felli}
\author[lnf,sapienza]{M. Ferrini}
\author[lnf]{M. Giardoni}
\author[lnf,sapienza]{T. Greci}
\author[lnf,sapienza]{A. Paolozzi}
\author[lnf]{L. Passamonti}
\author[lnf]{D. Piccolo}
\author[lnf]{D. Pierluigi}
\author[lnf]{A. Russo}
\author[lnf,sapienza,COR]{G. Saviano}
\author[napoli1]{S. Buontempo}
\author[napoli1,napoli2]{A. Cimmino}
\author[napoli1,napoli2]{M. de Gruttola}
\author[napoli1]{F Fabozzi}
\author[napoli1,napoli2]{A.O.M. Iorio}
\author[napoli1]{L. Lista}
\author[napoli1]{P. Paolucci}
\author[pavia]{P. Baesso}
\author[pavia]{G. Belli}
\author[pavia]{D. Pagano}
\author[pavia]{S.P. Ratti}
\author[pavia]{A. Vicini}
\author[pavia]{P. Vitulo}
\author[pavia]{C. Viviani}
%
\author[cern]{R. Guida}
\author[cern]{A. Sharma}
\address[lnf]{INFN Laboratori Nazionali di Frascati, Via E. Fermi 40, I-00044 Fr
ascati, Italy}
\address[sapienza]{Sapienza Universit\`a degli Studi di Roma
 "La Sapienza", Piazzale A. Moro, Roma, Italy}
\address[cern]{CERN CH-1211 Gen\'eve 23 F-01631 Switzerland}
\address[napoli1]{INFN Sezione di Napoli,
  Complesso Universitario di
Monte Sant'Angelo, edificio 6, 80126 Napoli, Italy }
\address[napoli2]{Universit\`a di Napoli Federico II, Complesso
Universitario di Monte Sant'Angelo, edificio 6, 80126 Napoli, Italy  }
\address[pavia]{INFN Sezione di Pavia and Universit\`a degli studi di
Pavia, Via Bassi 6, 27100 Pavia, Italy  }
\thanks[COR]{Corresponding author: Giovanna Saviano\\ E-mail address: 
giovanna.saviano@uniroma1.it}

\begin{abstract}
The CMS RPC muon detector utilizes a gas recirculation system called closed loop (CL) 
to cope with large gas mixture volumes and costs.
 A systematic study of CL gas purifiers has been carried out over 400 days between July 2008 and August 2009 
 at CERN in a low-radiation test area,  with the use of RPC chambers with currents monitoring,
 and gas analysis sampling points. The study aimed 
 to fully clarify the presence of pollutants, 
the  chemistry of purifiers used in the CL, and the regeneration procedure. 
Preliminary results on contaminants release and purifier characterization are reported. 
\end{abstract}

\begin{keyword}


RPC \sep CMS \sep gas \sep purifier \ detectors \ HEP \ muon
\end{keyword}

\end{frontmatter}



\section{Introduction}
The Resistive Plate Chamber (RPC) \cite{Santonico:1981sc}
 muon detector of the Compact Muon Solenoid (CMS) experiment\cite{:2008zzk}
utilizes a gas recirculation system called closed loop (CL) \cite{HAHNa}, \cite{HAHNb}
to cope with large gas mixture volumes and costs.
 A systematic study of Closed Loop gas purifiers has been carried out
  in 2008 and 2009 at the ISR experimental area of CERN  with the use 
  of RPC chambers exposed to cosmic rays with currents monitoring
 and gas analysis sampling points. Goals of the study 
 \cite{Abbrescia:2006LNF}
 were
 to observe the release of contaminants in correlation with the dark current increase
  in RPC detectors, to measure the purifier lifetime, to observe the presence of pollutants
   and to study the regeneration procedure.
Previous results had shown the presence of metallic contaminants,
 and an incomplete regeneration of purifiers
\cite{Saviano:2008Mumbai},\cite{Bianco:2009CMSNOTE}.
\par
The basic function of the CMS CL  system is to mix and purify the gas components 
in the appropriate proportions and to distribute the mixture to the individual chambers.
The gas mixture used is 95.2\% of C$_2$H$_2$F$_4$ in its environmental-friendly version R137a,
 4.5\% of $i$C$_4$H$_{10}$, and 0.3\% SF$_6$ to suppress streamers and operate in saturated avalanche mode.
 Gas mixture is humidified at the 45\% RH (Relative Humidity) level typically to balance ambient humidity,
which affects the resistivity of highly hygroscopic bakelite, and to improve efficiency
 at lower operating voltage. The CL is operated with a fraction of fresh mixture continuously injected into the
 system.
 Baseline design fresh mixture fraction for CMS is 2\%, the test CL system was operated at 10\% fresh mixture. The fresh mixture fraction is the fraction of the total gas content continuously replaced in the CL system with fresh mixture. The filter configuration is identical to the CMS experiment.
\section{Experimental setup and data sample}

In the CL system gas purity is guaranteed by a multistage purifier system:
   \begin{itemize}
       \item 
       The purifier-1 consisting of a cartridge filled 
       with 5\AA~ (10\%) and 3\AA~ (90\%) Type LINDE\cite{LINDE}
       molecular sieve\cite{GRACE} based on Zeolite manufactured by ZEOCHEM\cite{ZEOCHEM}; 
     \item
       The purifier-2 consisting of a cartridge filled with 50\% Cu-Zn 
       filter type R12 manufactured by BASF\cite{BASF} and
50\% Cu filter type R3-11G manufactured by BASF;
     \item
        The purifier-3 consisting of a cartridge filled with Ni 
	AlO$_3$ filter type 6525 manufactured by
	LEUNA\cite{LEUNA}. 
   \end{itemize} 
 \par
\begin{figure}[htbp]
  \begin{center}
    \resizebox{8cm}{!}{\includegraphics{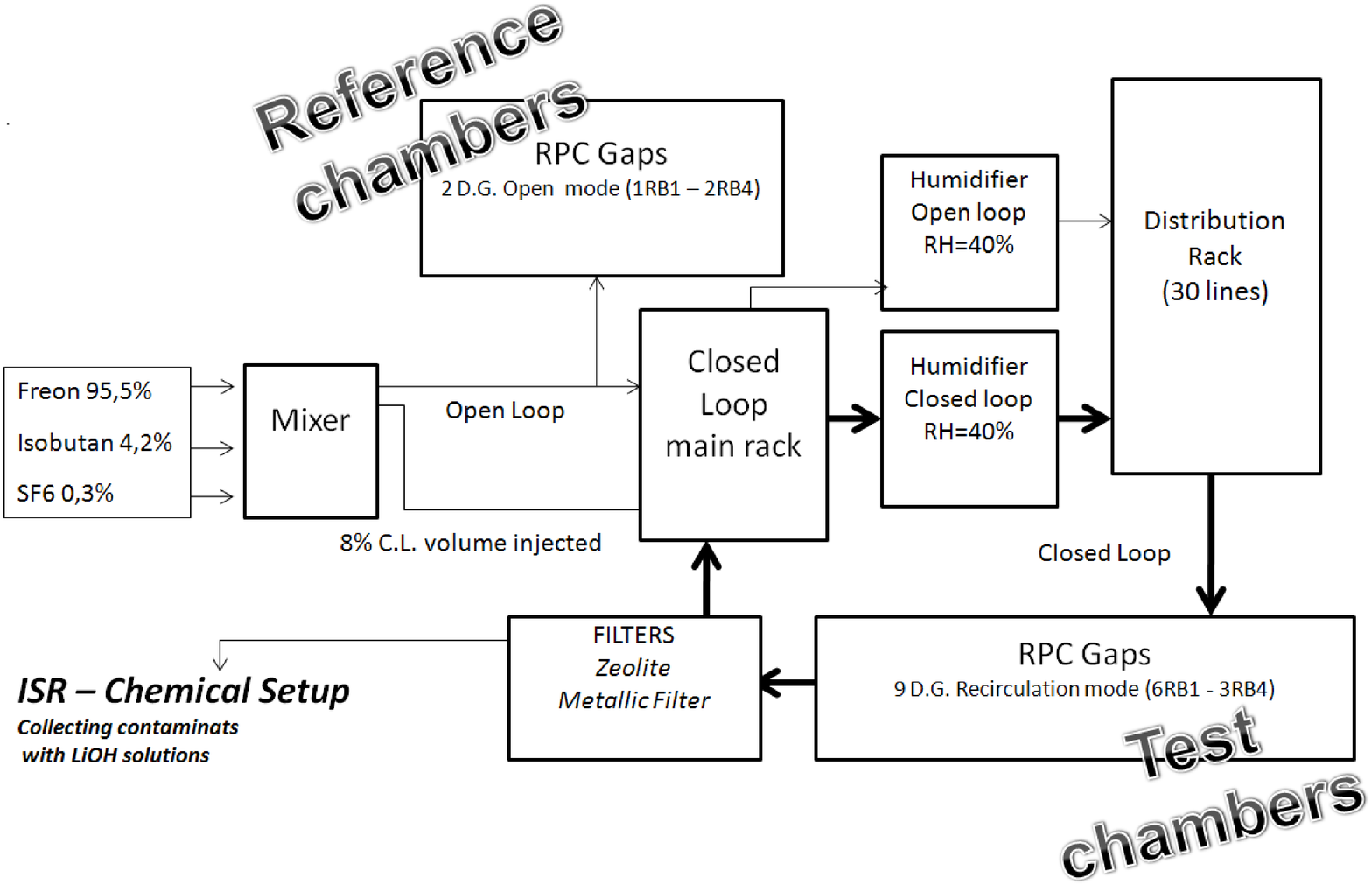}}
    \caption{Test setup with CL and open loop.}
 
    \label{FIG:setup}
  \end{center}
\end{figure}

The experimental setup  (Fig.~\ref{FIG:setup}) is composed of a CL 
system and an open mode gas system. A detailed description of the CL, the experimental setup, and the
filters studied can be found in \cite{Bianco:2009CMSNOTE}.
  The CL is composed of mixer, purifiers (in the subunit called ``filters`` in the Fig.~\ref{FIG:setup}) , recirculation
pump and distribution to the RPC detectors. 
 Eleven double-gap RPC detectors are installed, nine in CL and 
 two in open mode.
Each RPC detector has
two gaps (upstream and downstream) whose gas lines are serially
connected. The the gas flows first in the upstream gap and then in the downstream gap.
The  detectors are operated at a 9.2~kV power supply. 
The anode dark current drawn because of the high bakelite resistivity
is approximately 1-2~$\mu$A.
Gas sampling points before and after each filter in the
closed loop allow  gas sampling  for chemical and gaschromatograph
analysis. The system is located in a temperature and humidity 
controlled hut, with online monitoring of
environmental 
parameters.

\begin{figure}[htbp]
  \begin{center}
    \resizebox{9cm}{!}{\includegraphics{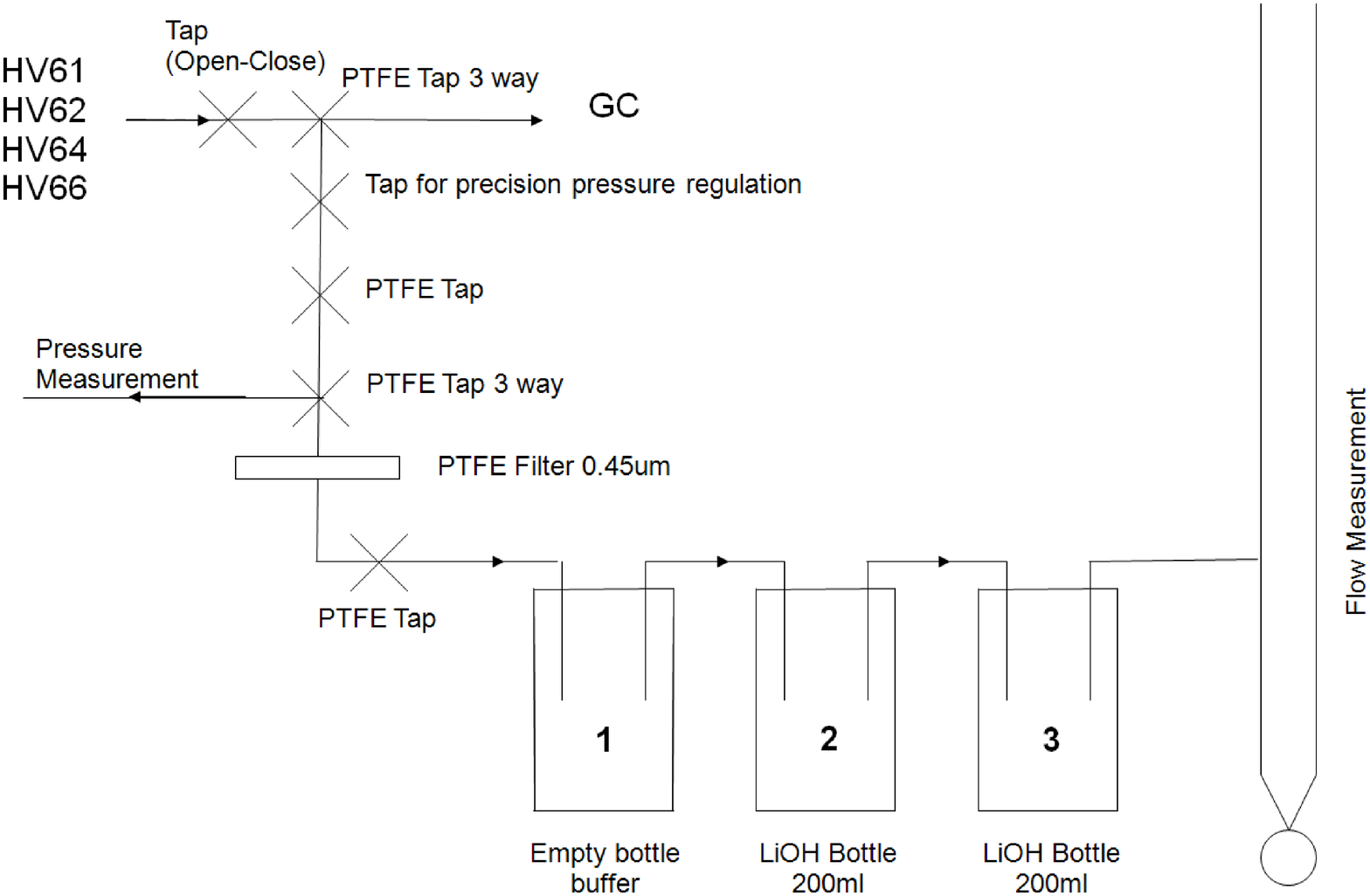}}
    \caption{Chemical setup}
    \label{FIG:chem_setup}
  \end{center}
\end{figure}

Chemical analyses have been performed in order to 
study the dynamical behaviour of dark currents increase in the  double-gap experimental setup and
correlate to the presence of contaminants,
measure lifetime of unused 
purifiers, 
 and identify contaminant(s) in correlation with the increase of  currents.
In the chemical analysis set-up (Fig.~\ref{FIG:chem_setup}) the gas is sampled
 before and after each CL purifier, and bubbled into a set of PVC flasks. The first flask is empty and
 acts as a buffer, the second and third flasks contain 250~ml solution of LiOH (0.001 mole/l
 corresponding to 0.024 g/l, optimized to keep the  pH of the solution at 11).
 The bubbling of gas mixture into the two flasks allows one to
 capture  a wide range of elements that are likely to be released by the system, such as 
  Ca, Na, K,  Cu, Zn, Cu, Ni, F.
At the end of each sampling line the flow is measured in order to have the total gas amount for
the whole period of sampling. Tygon filters ($0.45\mu$m) have been installed upstream  the flasks.
\par

\begin{figure}[htbp]
  \begin{center}
    \resizebox{9cm}{!}{\includegraphics{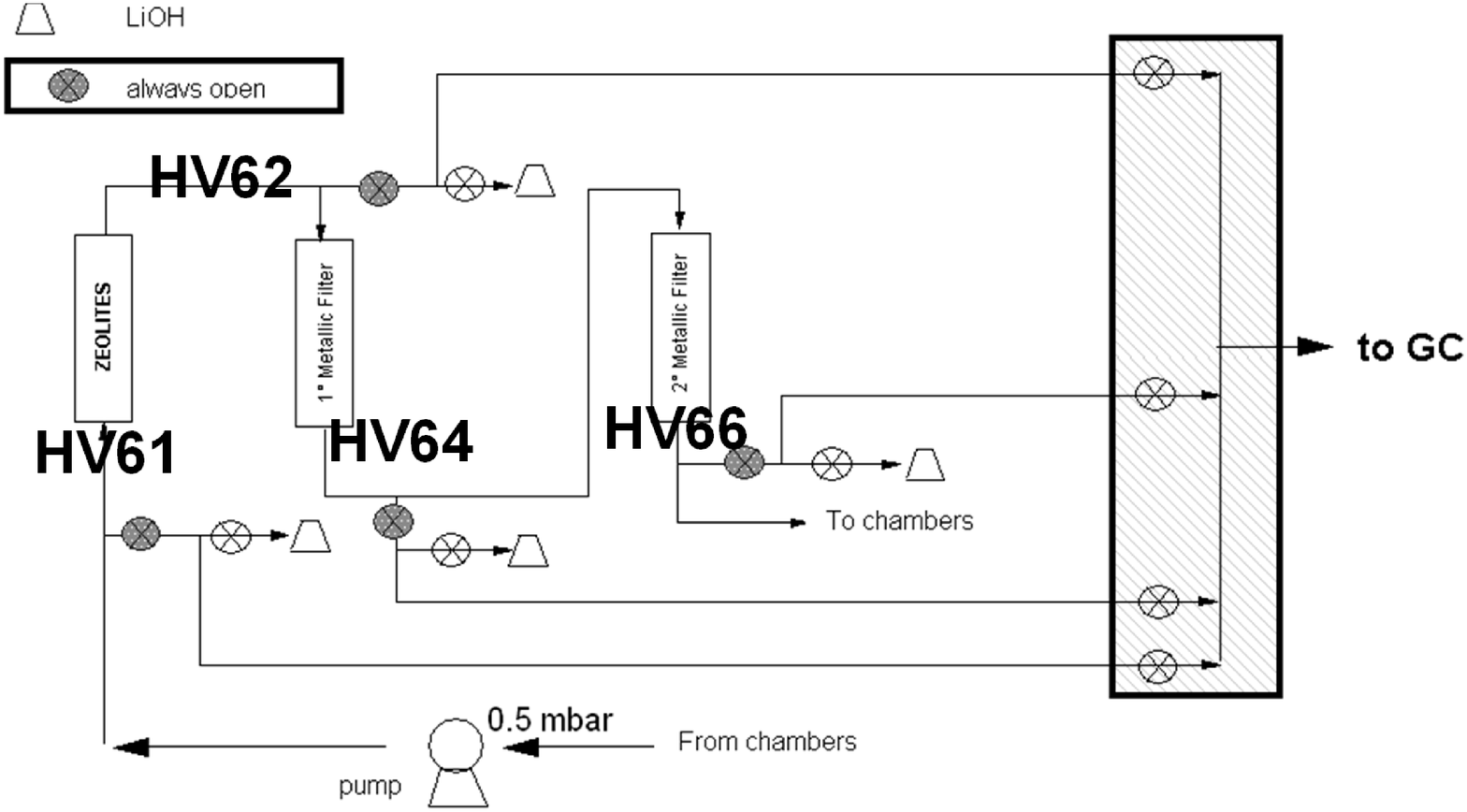}}
    \caption{Chemical setup sampling points}
    \label{FIG:sampling_point}
  \end{center}
\end{figure}

\begin{figure}[htbp]
  \begin{center}
    \resizebox{8cm}{!}{\includegraphics{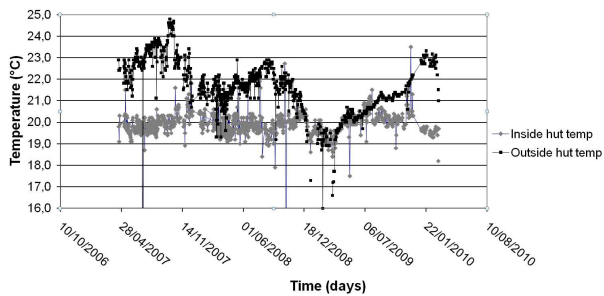}}
    \caption{Temperature distribution inside and outside the experimental hut.}
    \label{FIG:ISRtemp}
  \end{center}
\end{figure}

\begin{figure}[htbp]
  \begin{center}
    \resizebox{8cm}{!}{\includegraphics{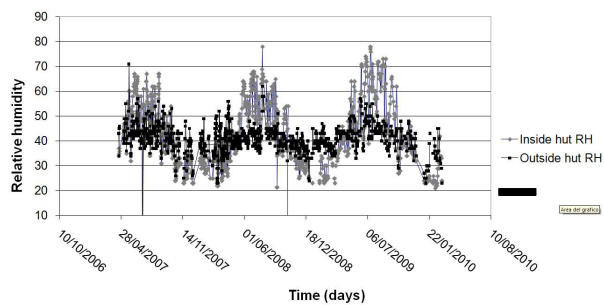}}
    \caption{Relative humidity inside and outside the experimental hut.}
    \label{FIG:ISRrh}
  \end{center}
\end{figure}

The sampling points (Fig.~\ref{FIG:sampling_point}) are located before the whole filters unit at position HV61,
 after purifier-1 (Zeolite) at HV62, after purifier-2 (Cu/Zn filter) at HV64 and after the Ni filter at position HV66.
RPC are very sensitive to environmental parameters (atmospheric pressure, humidity, temperature),
 this study has been performed in environmentally controlled hut with  
 pressure, temperature and relative humidity online monitoring.
 The comparison of temperature and humidity inside and outside the hut is displayed in Fig.~\ref{FIG:ISRtemp} and Fig.~\ref{FIG:ISRrh},
respectively, over the whole time range of the test. 
The inside temperature shows a variation of less than $\pm 0.5 \,^{\circ}{\rm C}$; 
the inside humidity still reveals seasonal structures between 35\% and 50\%, it is, however, 
much smaller than the variation outside.

 Gas mixture composition was monitored twice a day by gaschromatography, which also provided the
  amount of air contamination, stable over the entire data taking run and below 300 (100) ppm in closed (open) loop.
 Purifiers were operated with unused filter material.

\section{Results and discussion}

\begin{figure}[htbp]
  \begin{center}
    \resizebox{9cm}{!}{\includegraphics{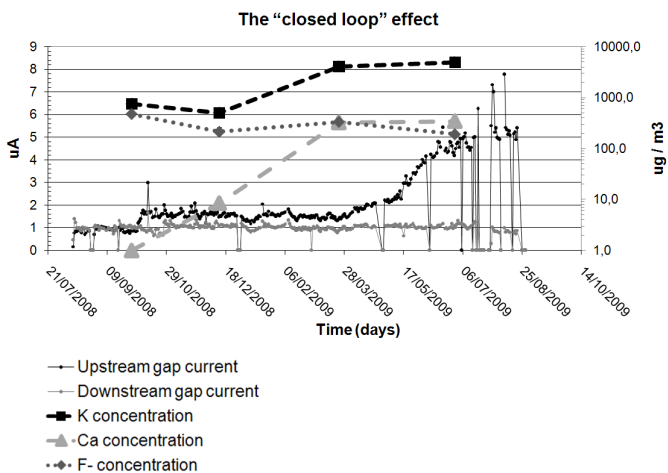}}
    \caption{Dark currents increase in the upstream gap and not in the downstream gap, correlated to the
    detection of contaminants in gas.}
    \label{FIG:closedloop}
  \end{center}
\end{figure}

\begin{figure}[htbp]
  \begin{center}
    \resizebox{8cm}{!}{\includegraphics{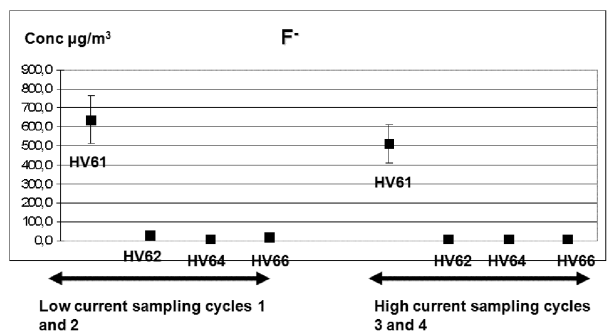}}
    \caption{Concentration of F$^-$ in gas as measured by the chemical setup. 
    Sampling point HV61 is before the zeolite
    purifier, the others after each purifier.}
    \label{FIG:fmeno}
  \end{center}
\end{figure}

The data-taking run was divided into cycles where different phenomena were expected. We have four cycles (Fig.~\ref{FIG:closedloop}), i.e., initial stable currents (cycles 1 and 2), 
 at the onset of the
 raise of currents (cycle 3), in the full increase of currents (cycle 4). Cycle 4 was terminated in order not to damage permanently
 the RPC detectors. The currents of all RPC detectors in open loop were found stable over the four cycles.
 Fig.~\ref{FIG:closedloop} shows the typical behaviour of one RPC detector in CL. While the current of the downstream
 gap is stable throughout the run, the current of the upstream gap starts increasing after about seven months. Such
 behaviour is suggestive of the formation of contaminants in the CL which are retained in the upstream gap, thus causing
 its current to increase, and leaving the downstream gap undisturbed. While the production of F$^-$ is constant during
 the run period, significant excess of K and Ca is found in the gas mixture in cycles 3 and 4. The production of F$^-$
  is efficiently depressed by the zeolite purifier (Fig.~\ref{FIG:fmeno}). The observed excess of K and Ca could be explained by a
  damaging effect of HF (continuously produced by the system) on the zeolite filter whose structure contains such elements.
  Further studies are in progress to confirm this model.

\section{Conclusions}
Preliminary results show that the lifetime of purifiers using unused material is approximately seven months. Contaminants
 (K, Ca) are released in the gas in correlation with the dark currents increase. The currents increase is observed only
 in the upstream gap. The study suggests that contaminants produced in the system stop in the upstream gap and affect its noise
 behaviour, leaving the downstream gap undisturbed. The presence of an excess production of K and Ca in coincidence with the
 currents increase suggests a damaging effect of HF produced in the system on the framework of zeolites which is based on K and
 Ca. Further studies are in progress to fully characterize the system over the four cycles from the
 physical and the chemical point of view. The main goal is to better schedule the operation and maintenance of filters for 
the CMS experiment, where for a safe and reliable operation the filter regeneration is presently performed several times per week.
A second run is being started with regenerated filter materials to measure
 their lifetime and confirm the observation of contaminants. Finally, studies in high-radiation environment at the CERN
 gamma irradiation facility are being planned.
\newline
\newline
{\bf Acknowledgements}
\newline
\newline
 The technical support of the CERN gas group
 is gratefully acknowledged. 
  Thanks are due to F.~Hahn  for discussions,
 and to  Nadeesha M. Wickramage, Yasser Assran  for help in data taking shifts.
 This research was supported in part by the Italian Istituto Nazionale
 di Fisica Nucleare and Ministero dell' Istruzione, Universit\`a e
 Ricerca.


\begin{thebibliography}{9}
\bibitem{Santonico:1981sc}
  R.~Santonico and R.~Cardarelli,
  ``Development Of Resistive Plate Counters,''
  Nucl.\ Instrum.\ Meth.\  {\bf 187} (1981) 377.
\bibitem{:2008zzk}
  CMS Collaboration,
  ``The CMS experiment at the CERN LHC'',
  JINST {\bf 3} (2008) S08004.
\bibitem{HAHNa} 
  M. Bosteels et al., ``CMS Gas System Proposal'', CMS Note 1999/018. 
\bibitem{HAHNb}
  L. Besset et al., ``Experimental
  Tests with a Standard Closed Loop Gas Circulation System'',    CMS
  Note 2000/040. 
\bibitem{Abbrescia:2006LNF}
   M.~Abbrescia {\it et al.}, ``Proposal for a Systematic Study of the CERN
   Closed Loop Gas System Used by the RPC Muon Detectors in CMS'',
   Frascati preprint LNF-06/27(IR), available at 
  {\tt  http://www.lnf.infn.it/sis/preprint/ }.
\bibitem{Saviano:2008Mumbai}
     G.~Saviano {\it et al.}, ``Materials studies for the RPC detector in CMS
     '', presented at the RPC07 Conference, Mumbai (India), January 2008.
\bibitem{Bianco:2009CMSNOTE}
  S.Bianco {\it et al.}, ``Chemical analyses of materials used in the
  CMS RPC muon detector'', CMS NOTE 2010/006.
\bibitem{ZEOCHEM} 
   Manufactured by ZEOCHEM, 8708 Uetikon (Switzerland).
\bibitem{BASF}
   BASF Technical Bulletin.
\bibitem{LEUNA}
  LEUNA Data Sheet September 9, 2003, Catalyst KL6526-T.   
\bibitem{GRACE}
  GRACE Davison  Molecular Sieves data sheet.
\bibitem{LINDE}
  LINDE Technical Bullettin.
\bibitem{Benussi:2008vs}
  L.~Benussi {\it et al.},
  ``Sensitivity and environmental response of the CMS RPC gas gain monitoring system,''
  JINST {\bf 4} (2009) 
  DOI:10.1088 1748-0221 4 08 P08006
  [arXiv:0812.1710 [physics.ins-det]].
\bibitem{Abbrescia:2006hk}
  M.~Abbrescia {\it et al.},
  ``HF Production In Cms-Resistive Plate Chambers,''
  Nucl.\ Phys.\ Proc.\ Suppl.\  {\bf 158} (2006) 30.  
  NUPHZ,158,30;
\bibitem{Aielli05}
 G.~Aielli {\it et al.}, 
 ``Fluoride production in RPCs
  operated with F-compound gases'',
   8th Workshop on Resistive Plate Chambers and Related Detectors,
  Seoul, Korea, 10-12 Oct 2005. Published in
  Nucl.Phys.Proc.Suppl. {\bf 158} (2006) 143.
\end{thebibliography}
\end{document}